\documentclass[cits]{PoS}
\usepackage{amssymb,fontenc,times,mathptmx,graphicx}
\usepackage{epstopdf}
\usepackage{slashed}

\title{Three easy exercises  in off-shell string-inspired methods }

\ShortTitle{Three easy exercises }

\author{\speaker{John Cornwall}\\
        University of California at Los Angeles (UCLA)\\
        E-mail: \email{cornwall@physics.ucla.edu}}

\abstract{Off-shell string-inspired methods (OSSIM) calculate off-shell QCD Green's functions using Schwinger-Feynman proper-time techniques, always in the background field method (BFM) Feynman gauge for technical convenience, and so far only at one loop.  We already know that these results are gauge-invariant, because this gauge realizes the prescriptions of the Pinch Technique (PT), a Feynman-graph formulation for any gauge, but the idea of the first exercise is to show this directly in OSSIM. In this exercise we extend proper-time OSSIM beyond the BFM Feynman gauge so that one can apply PT algorithms, and show that the intrinsic PT is equivalent to resolving ambiguities in OSSIM in other gauges.    In the second exercise we use forty-year-old rules of the author and Tiktopoulos for expressing loop integrals with numerator momenta directly in terms of Feynman parameters after momentum integration (the goal of OSSIM) and show that these rules elegantly and with economy of effort give rise, at least at one loop, to standard OSSIM  algorithms.  In the third exercise we apply world-line techniques to the problem of the breaking of adjoint strings, requiring a non-perturbative treatment that in the end reduces to a variant of the Schwinger result for production of $e^+-e^-$ pairs in an electric field.  This generalizes OSSIM to non-perturbative processes.}

\FullConference{From quarks and gluons to hadronic matter:  A bridge too far?\\
2-6 September, 2013\\
European Center for Theoretical Studies in Nuclear Physics and Related Areas (ECT*),
Villazzano, Trento (Italy)}

\begin{document}

\section{Introduction}

On-shell string-inspired techniques \cite{bk} are powerful algorithms to do the momentum-space integrals of   multi-leg multi-loop QCD S-matrix elements, yielding  simpler integrals over Feynman parameters.  Early on, Bern  and Dunbar \cite{berndun} realized that the power of the string-inspired methods for a $d=4$ NAGT relied only loosely on the actual machinery of string theory, which is designed to accommodate an infinity of states of higher and higher spin in critical dimensions such as 10 and 26.  In fact, the string-inspired methods could be found directly from field theory itself \cite{berndun,strass}, using as a basic tool Feynman-Schwinger proper-time methods. More recently off-shell string-inspired methods (OSSIM) have been used \cite{strass,cs} for elegant calculations of off-shell QCD amplitudes with proper-time methods.  Going off-shell raises the crucial issue of possible gauge dependence of these amplitudes.

  For reasons of technical simplicity OSSIM always use the background field method (BFM) Feynman gauge.  (We do not have space to discuss strictly string-theoretic arguments \cite{divecch} that a natural extension to off-shell of string-theoretic methods for on-shell field theory amplitudes does yield the BFM Feynman gauge.)   As it turns out, OSSIM amplitudes in this gauge are in fact gauge-invariant because they coincide with gauge-invariant amplitudes constructed with the principles of the Pinch Technique (PT) \cite{cornbinpap,corn076,corn099,papbin}, based on a rearrangement of Feynman graphs in an S-matrix element, so asking whether OSSIM amplitudes are gauge-invariant has already been answered, although indirectly.

Nevertheless, in our first exercise we investigate how to formulate OSSIM on its own terms (proper time) in an $R_{\xi}$ gauge, and how to apply the PT\footnote{It has been noted \cite{schubert} that the one-loop three-gluon vertex function calculated  by OSSIM in  the BFM Feynman gauge gives precisely the results of the PT \cite{corn099}.}.   In this connection, we can quote Feynman \cite{feyn} about his work on path-integral quantum mechanics:
\begin{quote}
There are, therefore, no fundamentally new results.  However, there is a pleasure in recognizing old things from a new point of view.
\end{quote}

In this paper, after a quick review of the PT, of OSSIM, and of their relationships, in the first exercise we extend OSSIM proper-time technology to an arbitrary gauge, and show that applying the so-called intrinsic PT is a way of removing ambiguities in the proper-time integrand for a general $R_{\xi}$ gauge.  In the second exercise, we use 40-year-old algorithms \cite{corn043} for expressing a Feynman graph with arbitrary momentum dependence in the numerator in terms of Feynman parameters after momentum integrations are done, and show that (at least at one loop) the results are as elegant and effective as OSSIM.  The third exercise shows how to use OSSIM to make gauge-invariant estimates of the tunneling probability for adjoint string breaking.

\section{A quick review of the Pinch Technique}

Although the PT is well-known we give a short review in order to set notation and emphasize certain similarities with OSSIM.  The claim of the PT is that with it one can reassemble parts of various off-shell Feynman graphs having different numbers of external legs (but the same number of loops) so that the result is a new $N$-point function that is completely gauge-invariant.   

\begin{figure}
\begin{center}
\includegraphics[width=5in]{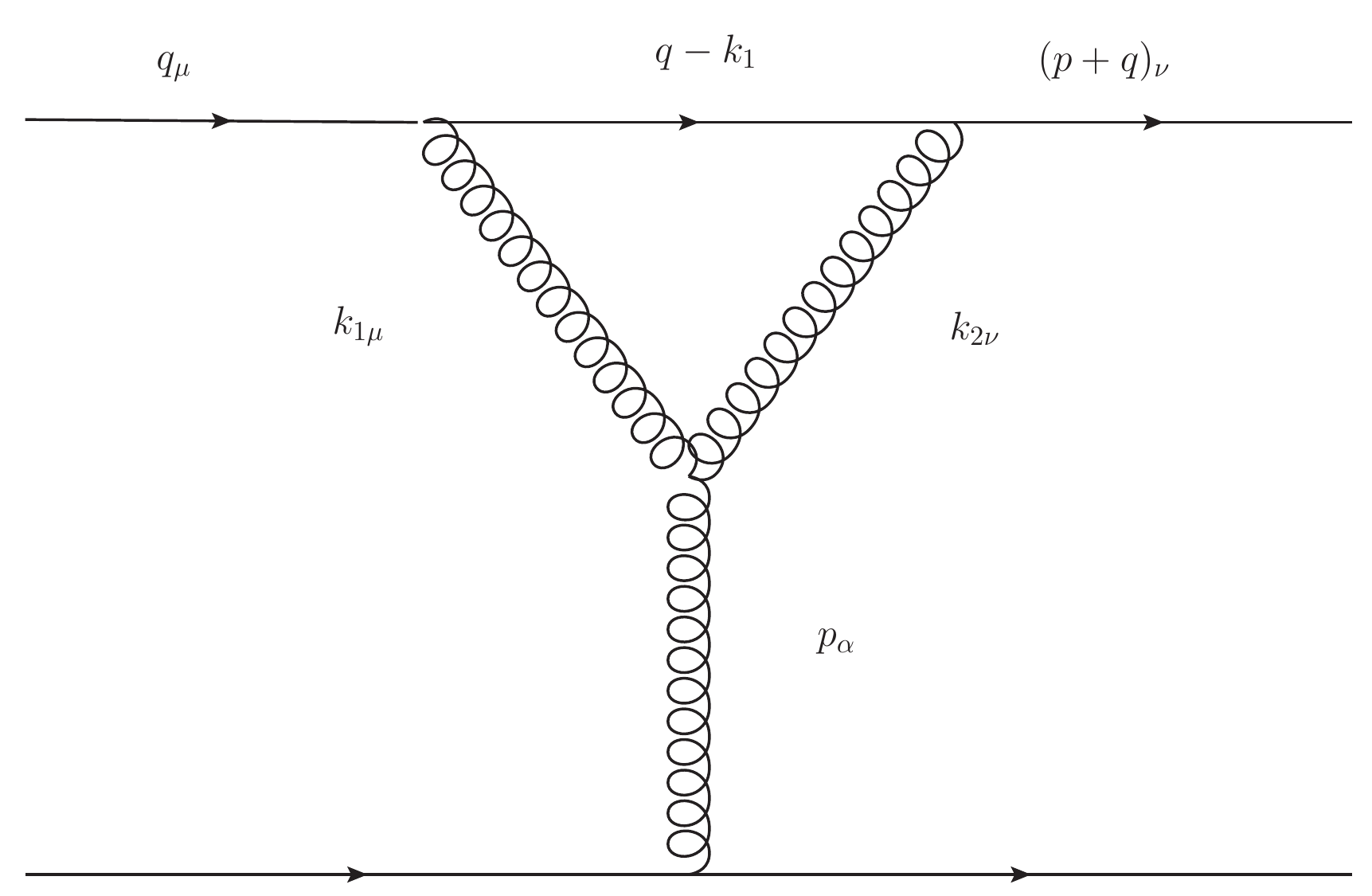}
\caption{A representative graph for illustrating the PT.}
\label{pt-graph}
\end{center}
\end{figure}

   Throughout this paper we work in Euclidean space.  To find, say, the one-gluon-loop proper PT self-energy \cite{corn076,cornbinpap} consider one of the graphs in the $qq\rightarrow qq$ S-matrix element of (Fig.~\ref{pt-graph})\footnote{For comparison with standard OSSIM notation it is convenient to choose $p,\;k_1$ as ingoing, but $k_2$ as outgoing at the three-gluon vertex.}.   Because it is part of an S-matrix element, all quarks are on-shell. 
 For future reference, we call the gluon not in a loop a {\em background (B)} gluon; the other two are {\em quantum gluons (Q)}.
For both the PT and OSSIM it is useful to decompose  the bare BQQ three-gluon vertex $\Gamma$ of Fig.~\ref{pt-graph}  into three parts, only two of which are used in OSSIM.  But the third part has all the complications that bring into question the gauge invariance of OSSIM.  In studying gauge invariance nothing is lost, but much simplicity is gained, by working only with the Feynman gauge, and we give the decomposition only for this special case; for the general case, see \cite {cornbinpap}.   
The three-gluon vertex in the figure is, aside from a group-theoretic factor:
\begin{equation}
\label{usualvert}
  \Gamma_{\alpha\mu\nu}(p,k_1,k_2)=(k_1+k_2)_{\alpha}\delta_{\mu\nu}-(k_2+p)_{\mu}\delta_{\alpha\nu}+(p-k_1)_{\nu}\delta_{\alpha\mu}.
\end{equation}
 Decompose it into convective, spin, and pinch terms:
\begin{eqnarray}
\label{vertdecompose}
\Gamma_{\alpha\mu\nu}(p,k_1,k_2) & = & \Gamma^C_{\mu\nu\alpha}+\Gamma^S_{\alpha\mu\nu}+\Gamma^P_{\alpha\mu\nu } \\ \nonumber
\Gamma^C_{\alpha\mu\nu}  & =  & (k_1+k_2)_{\alpha}\delta_{\mu\nu},\;\Gamma^S_{\alpha\mu\nu}=-2p_{\mu}\delta_{\nu\alpha}+2p_{\nu}\delta_{\mu\alpha}\\ \nonumber
 \Gamma^P_{\alpha\mu\nu } & = & -k_{2\nu}\delta_{\mu\alpha}-k_{1\mu}\delta_{\nu\alpha}.
\end{eqnarray} 
Here $\Gamma^C$ is the convective vertex, independent of the quantum-field spin; $\Gamma^S$ is the spin generator  of the quantum field (akin to $\sigma_{\mu\nu}/2$ for fermions); and $\Gamma^P$ is the pinch  part of the vertex that triggers the Ward identities defining the PT.  It, and only it, has   longitudinal momenta in it.  
In the BFM Feynman gauge, the BQQ bare vertex coupling two quantum gluons to one background gluon is just $\Gamma^F\equiv \Gamma^C+\Gamma^S$; there is no pinch part.  This is why the BFM Feynman gauge is special for the PT:  No pinches occur.  Moreover, on the B line the vertex $\Gamma^F$ obeys a QED-like Ward identity.
(Note that this gauge has   Feynman rules for ghosts and four-gluon vertices that differ from those of the non-background Feynman gauge.)

\begin{figure}
\begin{center}
\includegraphics[width=2in]{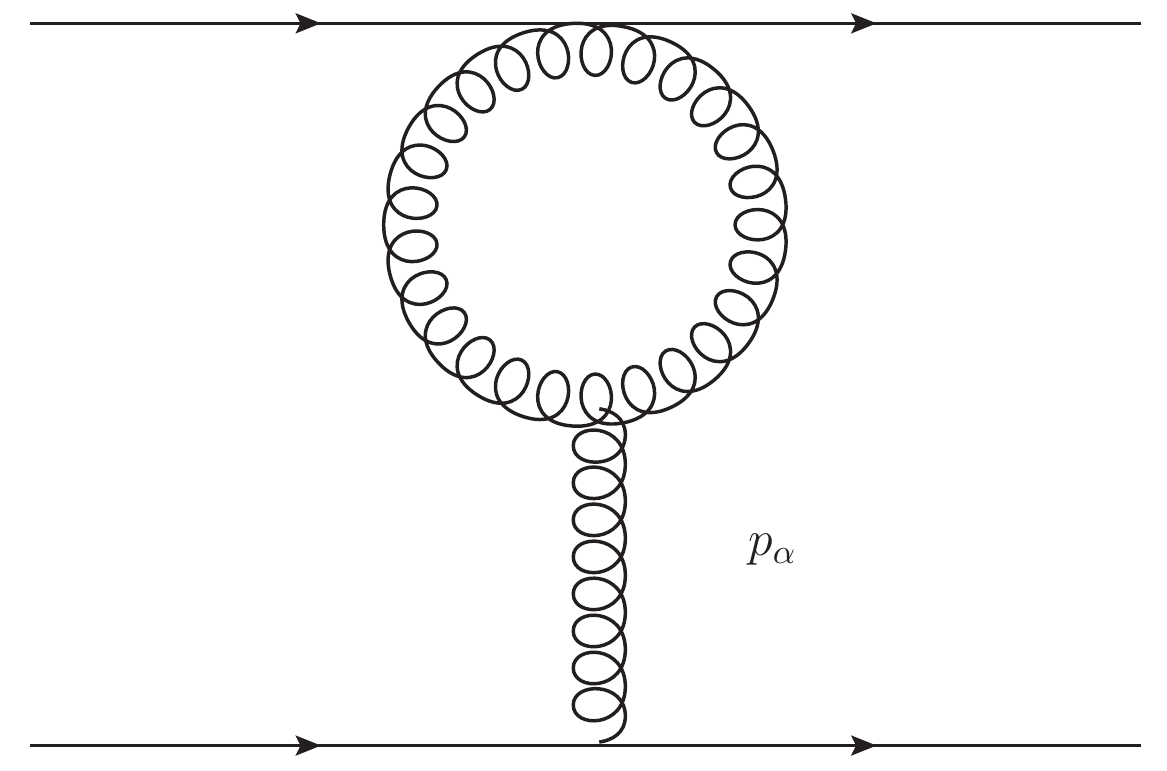}
\caption{\label{pinch-2}    Pinch parts come from longitudinal gluon momenta triggering Ward identities that replace quark lines by $\pm 1$.   }
\end{center}
\end{figure}

The pinch vertex parts of Eq.~(\ref{vertdecompose} generate  Slavnov-Taylor identities.  For Fig.~\ref{pt-graph} this reduces to  simple Ward identities such as:
\begin{equation}
\label{quarkwid}
k_{1\mu}\gamma_{\mu}=S^{-1}(q+k_1)-S^{-1}(q)\rightarrow  S^{-1}(q+k_1);
\end{equation}
the on-shell inverse propagator vanishes.  The remaining propagator cancels (pinches out) the propagator labeled $q+k_1$, leaving only a factor 1, so that the pinch part of Fig.~\ref{pt-graph} yields a graph with the topology shown in Fig.~\ref{pinch-2}.  In general, pinch parts  replace internal quark lines by $\pm 1$.  Note that these former contributions to a three-point vertex have become propagator parts; interpreted as proper self-energy parts, they have a factor of an inverse gluon propagator $\Delta^{-1}(p_1+p_2)$.
That is, to define a proper (one-particle irreducible, or 1PI) gluon self-energy, one needs to insert the unit product $1=\Delta^{-1}\Delta$ of a gluon inverse propagator and the propagator between the new vertex induced by pinching and the quark vertex, and associate the inverse propagator with the pinch part.  In fact, all pinch parts carry this inverse propagator factor, which is the basis of the intrinsic pinch technique.
It is easy to check that the group-theoretic coefficient, ignored so far, always involves a commutator
$[T_a,T_b]$ of the generators of whatever group representation the external on-shell particles carry.

The final result of these manipulations   is that the sum of the conventional self-energy parts   plus the pinch parts of Fig.~\ref{pinch-2}  yields a gauge-invariant   1PI BB proper gluon self-energy.  

Both the PT and OSSIM use polarization vectors for {\bf off-shell} background gluons  (the gluon of momentum $p$ in Fig.~\ref{pt-graph}), satisfying $\epsilon (p)\cdot p=0$ even though $p^2\neq 0$ .  
   Such a background gluon of momentum $p$ couples to
$ \bar{u}(p+q)\gamma_{\mu}u(q)\equiv \epsilon_{\mu}(p;q)$
where the spinors are on-shell.  (One is reminded of the fermion-antifermion polarization vectors of on-shell algorithms \cite{berends,xu}.)
After pinching (which removes internal quark lines), this type of polarization vector is generated for Fig.~\ref{pt-graph}.
   That the polarization vectors are transverse off-shell is an important point, since it is a source of ambiguities, related to application of the intrinsic PT, that we discuss in Sec.~\ref{ex1}.  It is also important for OSSIM, where it occurs naturally.  Note that spacelike (Euclidean) momenta for the background gluons are assured by using equal-mass Minkowski-space spinors.

 A final remark: Ward identities in the PT are generically integrals of total derivatives in  Feynman parameters, as illustrated in \cite{corn141}. In Sec.~\ref{ossimrev} we show that the action of a longitudinal gauge-boson momentum triggers total derivatives in proper times.  Since one can always choose Feynman parameters to be scaled proper times, the field-theoretic and the OSSIM approach give the same results.  Ward identities are important in the PT because they trigger the pinch parts that are added to (or subtracted from, depending on the point of view) conventional Feynman graphs.  In OSSIM certain proper-time derivatives occur (see Eq.~(\ref{strassres})) that are integrated by parts, and called IBP terms.  IBP parts and pinch parts should not be confused; they are not the same, in general.

\section{\label{ossimrev}  A brief review of the field-theoretic basis of OSSIM}

   So far, OSSIM have been applied to one-loop background-field actions with arbitrarily many off-shell background gluon lines attached, and this is all we will consider.  OSSIM   give expressions for off-shell Feynman graphs after momentum-space integration that is equivalent to an overall proper-time integral.  The result is an effective action depending on a specific string-inspired choice of Feynman parameters.    This effective action is a sum of terms (including ghosts) of the form
\begin{equation}
\label{effactform}
\Gamma_S\{B_{\mu}\}=\frac{-1}{2}{\rm Tr}\log \Delta^{-1}.
\end{equation}
The coefficient -1/2 refers to a single scalar; there are differing prefactors for fermions, ghosts, charged scalars, and gluons.
Here $\Delta$ is the one-loop propagator of the {\em quantum} field $Q_{\mu}$, propagating in the presence of {\em background fields} $B_{\mu}$.

Write the proper-time loop propagator for a charged scalar of mass $m$ in the presence of the background field:    
\begin{equation}
\label{scalprop}
\Delta (x-y)=\textrm{P}\int_0^{\infty}\mathrm{d}se^{-m^2s}\int_x^y \{\mathrm{d}z_{\mu}\}
\exp \{-\int_0^s \mathrm{d}\tau[\frac{\dot{z}_{\mu}^2}{4}-ig\dot{z}\cdot B_{\mu}(z)]\};
\end{equation}
to form the logarithm, divide the integrand by $s$.  The integration over $z$ is functional, and the dot indicates a proper time derivative. The trace operation in the effective action includes the prescription that $x=y$, and the functional integration is over a closed worldline.  The Wilson-loop gauge interaction gives the convective part of the gauge vertex, no matter what spin is in the quantum loop, and in particular it gives $\Gamma^C$ of Eqn.~(\ref{vertdecompose}) for any field in the loop carrying gauge charge.  

The background gauge potential is a sum of terms:
\begin{equation}
\label{nagtpot}
B_{\mu}(z)=\sum_{i=1}^N t_{a_i}\epsilon_{i\mu}(p_i)e^{ip_i\cdot z}
\end{equation}
and we seek the $\mathcal{O}(g^N)$ term in the effective action that has each background gluon polarization vector $\epsilon_i$ exactly once.  The corresponding current for the world-line variable is:
\begin{equation}
\label{current}
K_{\mu}(\tau )=\sum^N \delta (s - \tau_i)(\epsilon_{i\mu}\partial_{\tau_i}+ip_{i\mu})
\end{equation}  
where the plane waves act at proper times $\tau_i$.  It is discontinuous because of the momentum transfer at each vertex, but 
in the limit $N\rightarrow \infty,\;k_{i+1}-k_i=\mathcal{O}(1/N)$, the current is smooth.  We have written the current for one ordering of the color generators; the whole result is a sum over all orderings, modulo cyclic permutations.

\begin{figure}[ht!]
\begin{center}
\includegraphics[width=2in]{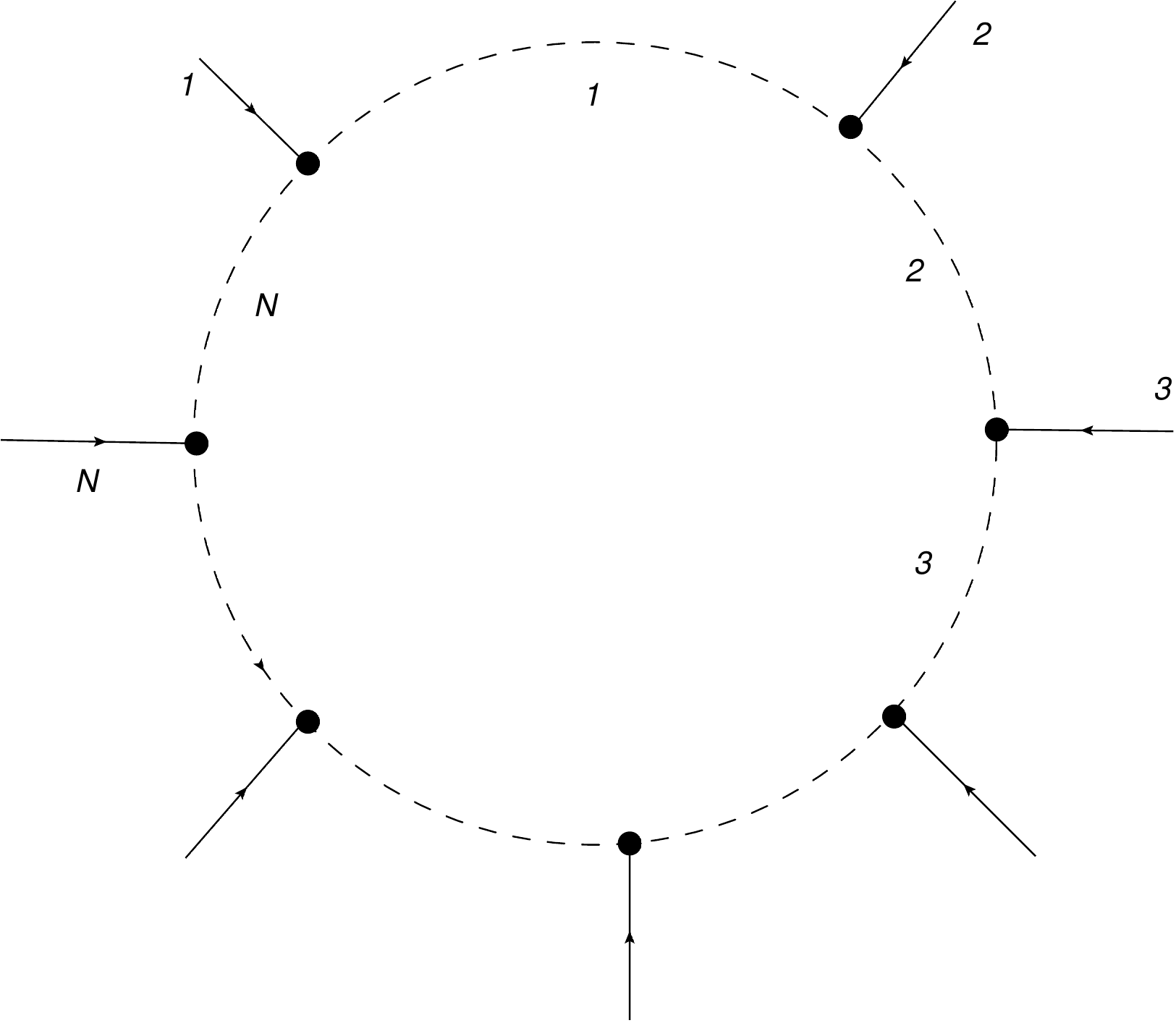}
\caption{\label{1loop}   A one-loop   graph.  The numbers label incoming momenta (external, or background, lines) and Feynman parameters on the dashed (internal, or quantum) lines.}
\end{center}
\end{figure}

It turns out that the scaled proper times $x_i\equiv \tau_i/s$ correspond to a choice of Feynman parameters for the effective action expressed as a Feynman graph.  
This equivalence is generally of the form shown in Fig.~\ref{1loop}.
Strassler \cite{strass} gives an example for a scalar loop attached to $N$ NAGT background gluons of momenta $p_i$, color matrices $t_{a_i}$, and polarizations $\epsilon_{i\mu}$.  After the path integration called for in Eqn.~(\ref{scalprop}) one finds  (with a slight change in notation and ignoring UV regulation) for the effective action corresponding to a single cyclic permutation of color labels:
$$\Gamma \{B_{\mu_i}\} = \frac{(ig^2)^N\mathrm{Tr}(t_{a_N}\dots t_{a_1} )}{16\pi^2}\times$$
\begin{equation}
\label{strassres} 
\times\int_0^{\infty}\frac{ds}{s^{3-N}}\int\![\mathrm{d}x_{ij}]
\exp [s\sum_{i<j}p_i\cdot p_j G_B (x_{ij})] \exp [\sum_{i<j}(-i(p_i\cdot \epsilon_j-p_j\cdot
\epsilon_i)\dot{G}_B(x_{ij})+\epsilon_i\cdot \epsilon_j\ddot{G}_B(x_{ij})]
\end{equation}
where it is understood that only the term multilinear in the $\epsilon_i$ is to be saved.  Here the sums over $i,j$ run from 1 to $N$.  The gluons are not on shell, and their polarization vectors are simply counting devices.  However, it turns out that terms involving $\epsilon_i\cdot p_i$ do not contribute, so the polarization vectors are effectively transverse (just as for the PT).  In this integral the $G_B$ are bosonic Green's functions, given in Eqn.~(\ref{bosgreen}), and we define
\begin{equation}
\label{feynpar}
[\mathrm{d}x_{ij}]=\prod \{\int_0^1\mathrm{dx_i}\}\delta (1-\sum x_i).
\end{equation}
The $x_i$ are linear combinations of conventional Feynman parameters (see the Appendix), with $x_{ij}\equiv x_i-x_j$ (see the Appendix).  At the same time, they are proper times scaled by the overall proper time $s$:  $x_i=\tau_i/s$.  The bosonic Green's function is
\begin{equation}
\label{bosgreen}
G_B(x_{ij})=|x_{ij}|-(x_{ij})^2,
\end{equation}
and the overdots indicate derivatives.  
  The Bern-Kosower prescription, followed by essentially all OSSIM practitioners, is to integrate the double-derivative term by parts, ignoring the surface terms---hence their acronym IBP, for integration by parts.

The power of OSSIM are only seen when higher spins in the loop are considered.  It is here where the choice of gauge enters, according to the form of the quantum gluon propagator.  For OSSIM this form is:
\begin{equation}
\label{quantprop}
\Delta^{-1} = D^2-\Sigma_{\mu\nu}G^{\mu\nu}
\end{equation}
where $D_{\mu}+iB_{\mu}$ is the covariant propagator of the quantum field in the presence of the background field, $\Sigma_{\mu\nu}^{(\alpha\beta )}$ is the spin matrix for the spin of the quantum field, $\alpha ,\beta$ label the group representation of the spin (these indices are contracted with those of the propagator), and $G_{\mu\nu}$ is the field-strength tensor of the background field.     For a spinor loop  $\Sigma_{\mu\nu}$   would be $(1/2)\sigma_{\mu\nu}$, and for a gluon  the simplest form of the spin generator is
\begin{equation}
\label{gluspin}
\Sigma^{(\alpha\beta)}_{\mu\nu}=i[\delta^{\alpha}_{\mu}\delta^{\beta}_{\nu}-
\delta^{\beta}_{\mu}\delta^{\alpha}_{\nu}].
\end{equation}
   Spins are incorporated into the proper-time integral through integration over Grassman coordinates, as is well-known.

Of course, for a gluon the free propagator of Eq.~(\ref{quantprop}) is in a Feynman gauge;  in fact, it is the BFM Feynman gauge.  The contribution of   $\Sigma^{(\alpha\beta)}_{\mu\nu}$ to the BQQ vertex is precisely that of the vertex $\Gamma^S$, and there is no pinch vertex $\Gamma^P$.   So   OSSIM Green's functions are chosen to be in BFM Feynman gauge, and are thereby gauge-invariant.

In OSSIM it is not necessary to work in the BFM Feynman gauge, but if one does then there are no pinch terms, either from the (Feynman-gauge) propagator or from the vertex $\Gamma^F$.  In any other gauge, such as the BFM $R_{\xi}$ gauge, there  are pinch terms.   We now incorporate pinch vertices into OSSIM, and note that they generate ambiguities.  It appears that their resolution is equivalent to the intrinsic PT \cite{corn099,cornbinpap}.

\section{\label{ex1}  Exercise 1:  OSSIM in an arbitrary covariant gauge:  Resolving ambiguities}

The convective vertex 
\begin{equation}
\label{convert} 
g\dot{z}\cdot B_{\mu}(z)
\end{equation}
corresponds in coordinate space to $ i  \stackrel{\leftrightarrow}{\partial_{\alpha}}$; the derivatives act on the quantum-field propagator segments.  With momentum labels as in Fig.~\ref{pt-graph} this is $(k_1+k_2)_{\alpha}$.  To go to an arbitrary gauge requires adding pinch vertices $\Gamma^P$  (see Eq.~(\ref{vertdecompose})).  These too are propagator derivatives:
\begin{equation}
\label{propder}
\partial_{1\alpha}\Delta_{F0}(x_1-x_2) = \langle -\frac{\dot{z}_{\alpha}}{2}(\tau_1=0;x_1,x_2,s)\rangle = -ik_1.
\end{equation}
But from the point of view of a proper-time integral $\dot{z}_{\alpha}(\tau_1=0)$ is ambiguous, because this proper velocity could refer either to $k_1$ or to $k_2$.  One should specify which propagator is being differentiated by a choice of $\tau_1=0\pm \epsilon$.  The convective vertex resolves this ambiguity by using the average, a familiar choice from Fourier transforms.

The same ambiguity appears in Ward identities.    The rules for both OSSIM and the PT say that background longitudinal terms such as $p_{\alpha}$ can be dropped in any vertex, because the polarization vector to which it is attached is transverse.   But suppose we use momentum conservation ($p=k_2-k_1$) to change the convective vertex as follows:
\begin{equation}
\label{changeconv}
(k_1+k_2)_{\alpha}=(2k_1+p)_{\alpha}\rightarrow 2k_{1\alpha}\;\textrm{or}\;(2k_2-p)_{\alpha}\rightarrow 2k_{2\alpha}.
\end{equation}
Now the Ward identity 
\begin{equation}
\label{badwid}
p\cdot (k_1+k_2)=k_2^2-k_1^2
\end{equation}
is changed to, {\em e. g.}: 
\begin{equation}
\label{newbadwid}
2p\cdot k_1\equiv k_2^2-k_1^2+p^2
\end{equation}
which is correct only if $p^2=0$.  In Euclidean space this mass-shell condition means $p_{\alpha}=0$ so that $k_{1\alpha}=k_{2\alpha}$, and there is no discontinuity.  To avoid off-shell ambiguity we must use the standard convective vertex $(k_1+k_2)_{\alpha}$.

All this is closely-related to the intrinsic PT, which begins with the usual Feynman graphs for, say, the three-gluon vertex and drops those terms that have inverse background-field propagators.  Such terms, and many others, arise from the Slavnov-Taylor identities generated by the pinch vertices $\Gamma^P$. A typical term for the vertex in Fig.~\ref{pt-graph} is:
\begin{equation}
\label{stident}
 -k_1\cdot (k_2+p)\delta_{\nu\alpha}+\dots = (p-k_2)\cdot (p+k_2)\delta_{\nu\alpha}+\dots = (p^2-k_2^2)\delta_{\nu\alpha}+\dots.
\end{equation} 
In the intrinsic PT, the term in $p^2$ should be dropped, because this term is the inverse propagator of a background gluon.  This is related to the principle of removing ambiguity by averaging: In the term $k_1\cdot k_2$ in the original form of the ST identity, we should make the replacement:
\begin{equation}
\label{replace}
\dot{z}(\tau_1-\epsilon )\cdot \dot{z}(\tau_1+\epsilon )\rightarrow \frac{1}{2}[\dot{z}(\tau_1-\epsilon )^2+
\dot{z}(\tau_1+\epsilon )^2]. 
\end{equation} 
This is tantamount to dropping the $p^2$ in the identity
\begin{equation}
\label{momcons}
k_1\cdot k_2=\frac{1}{2}(k_1^2+k_2^2-p^2);
\end{equation}
precisely the requirement of the intrinsic PT.  

Perhaps behind these simple remarks there is a principle for the PT that goes beyond the standard manipulation of Feynman diagrams.

\section{Exercise 2:  OSSIM results through old algorithms} 
 
   We use here a set of rules formulated long ago \cite{corn043} for giving the numerators of Feynman graphs in terms of Feynman parameters {\em after} momentum integrations are done.  At least for one-loop graphs with arbitrarily many background gluons attached, these old rules, based on graph topology, map directly onto OSSIM  where numerators are written in terms of bosonic and fermionic Green's functions on a closed string.  But these older methods can go much further, by dealing easily with graphs in a general $R_{\xi}$ gauge.  An example would be  the three-gluon proper vertex \cite{corn099,cornbinpap}.

The rules are a straightforward extension of much older rules based on graph topology for finding the form of the denominator $D$ of any Feynman graph, in terms of Feynman parameters and after all  momentum integrations.  The algorithms are too complicated (since they apply to an arbitrary Feynman graph) to give here. They involve, for example, finding all the chord sets of a graph, defined as a set of lines whose complement is a tree graph, and cut sets, lines whose complement is two disjoint trees, and forming sums of products of corresponding Feynman parameters.
These rules simplify enormously for a one-loop graph, and  have a particularly elegant form in OSSIM, where the Feynman parameters used are not those that would naturally be connected with, say, the graph of Fig.~\ref{1loop}.  The natural choice is to assign a Feynman parameter $\alpha_i$ with each line as labeled in the figure.  But the string-inspired choice is to use the variables $x_i$ described in the Appendix.  (In string theory, the $x_i$ are essentially Koba-Nielsen variables specifying the location of external vertices on the world sheet.)  

We give one example of the use of these algorithms.  In the numerator before momentum integration, a term linear in the loop momentum $k$ is replaced by its shifted value:
\begin{equation}
\label{kshift}
k \rightarrow \sum_j x_{ij}p_j.
\end{equation}
The power of this expression is that it holds for all $i$ (by momentum conservation).  Terms of quadratic and higher orders in $k$ involve pairings that are algorithmically just like Wick contractions.  For one-loop graphs with a relatively small number of lines the pairing coefficients are trivial to calculate.
One should note also that the spin vertex $\Gamma^S$ has no numerator momenta, which considerably simplifies the treatment of these vertices.

Finally, for one-loop graphs explicit calculation shows that it is no more work to apply the algorithms to a one-loop graph than it is to apply standard OSSIM.  This work simply amounts to rewriting the OSSIM results in a more useful form.  We hope these old algorithms can be revisited in much more detail and that they will prove useful.

 \section{Exercise 3:  OSSIM and adjoint string breaking}

The adjoint string can break,   an effect that both depends on the gluon mass $m$ and is strongly connected with the meaning of this mass.   This effect is missing from conventional Schwinger-Dyson equation studies (as reviewed in \cite{cornbinpap}), where there is no adjoint string to break. On the lattice, interplay between string breaking and the mass is obscured by finite-size lattice effects, the running of $m$ with momentum, and the fact that the Minkowski-space regime is inaccessible.  

Fundamental-string breaking was studied long ago \cite{nussinov} as a simple rewriting of the Schwinger instability ($e^+-e^-$ pair production in a constant electric field) for quarks with a constituent mass, but it is not quite so obvious how to apply these results to adjoint string breaking.  Clearly there is string breaking via unstable pair production\footnote{Not to be confused with the Nielsen-Olesen \cite{nielsen} magnetic instability, which is cured by a gluon mass.}, and it has to be described gauge-invariantly.  We recognize that this calls for the calculation to be done in the BFM Feynman gauge, and in OSSIM terms it amounts to replacing the discontinuous sum of perturbative gluons used in Eq.~(\ref{vertdecompose}) by a single space-time-dependent background potential.  
Earlier work\cite{corn080,corn121} provided a complementary view, in which the breakable adjoint string is produced by Wilson-loop perimeter (not area) terms for center vortices whose transverse extent is governed by the gluon mass $m$.  In this view, the static adjoint potential rises essentially linearly and then breaks at a distance where the adjoint string $\sigma_A$  has stored up an energy of about  $2m$.  

An effective and accurate way \cite{corn141} to account for the gluon mass at  low energies is just to replace the massless gluon propagator of Eq.~(\ref{quantprop}) with a massive one ``by hand":
\begin{equation}
\label{massprop}
\Delta^{-1}  = D^2 + m^2-\Sigma_{\mu\nu}B^{\mu\nu}.
\end{equation}
This works because (as long as $m$ is approximated by a constant, non-running mass) the necessary Ward identities are preserved, as well as certain supersymmetry relations \cite{brod} among three-vertices that follow from the PT, provided that the spin 0, 1/2, and 1 particles in the loop are given the same mass.  A paper in progress by the author will provide details of the accuracy of the approximation in Eq.~(\ref{massprop}).

Next, we will replace the unbroken adjoint string by a  chromoelectric field $\mathcal{E}$ that is approximately constant over a transverse distance $1/m$.   To estimate the parameters $\mathcal{E},\;m$ and zero-momentum coupling $g$ that enter the rate of adjoint string breaking we turn to   an extended gluon-chain model reported \cite{corn139} at the first Trento QCD-TNT Workshop.    The model hypothesizes a duality beteween the number density per unit area $\rho_M$ of chromomagnetic center vortices piercing an area and the   number density per unit area $\rho_E$ of ordinary (chromoelectric) gluons in the gluon-chain area. If the average  separation between gluons in this area is $\zeta$, we have
\begin{equation}
\label{dual}
\rho_E=\frac{1}{\zeta^2}= \textrm{(by duality)} \;\rho_M.
\end{equation}
 where $\sigma_F$ is the fundamental string tension.  An ancient model for this string tension\footnote{There is a typo in Ref.~\cite{corn139}; on p.~7 replace $\zeta$ by $1/\zeta$.} gives $\sigma_F \approx 2\rho_M = 2/\zeta^2$. Another relation for $\sigma_F$ equates the energy of a gluon chain of length $R$, which is $\sigma_FR$, to $R$ times the linear energy density of gluons, or $mR/\zeta$. This yields $\zeta \approx 2/m$ and $\sigma_F\approx m^2/2$.   For the adjoint string the electric flux equals the coupling:
\begin{equation}
\label{flux}
(\pi/m^2)\mathcal{E} = g,
\end{equation}
 and the adjoint string tension is the chromoelectric energy per unit length, or
\begin{equation}
\label{sigmaa}
\sigma_A= [\frac{\mathcal{E}^2}{2}][\frac{\pi}{m^2}] \approx m^2
\end{equation}
where we use approximate Casimir scaling:  The adjoint string tension $\sigma_A$ is about  $2\sigma_F$.
Combining these relations yields $g^2/(4\pi)\approx 1/2$, a   satisfactory value  phenomenologically.  A mass
$m\approx$ 600 MeV yields $\sigma_F\approx$ 0.18 GeV$^2$, as observed.  The minimum string-breaking length $\ell$   is about $2m/\sigma_A\approx 2/m\approx$  0.7 Fm; transverse momenta of the produced pair may double this.

Of course, one need not invoke the model to take on purely phenomenological grounds the relations $g^2/(4\pi) = 1/2,\;\sigma_F=m^2/2$, as we will use in our final formula.

Next,   the gluon propagator of Eq.~(\ref{massprop})  has a spin-1 generator $\Sigma_{\mu\nu}$, not spin-1/2 as for Schwinger.  Supplying a spin-color factor for $N_c$ colors, we get the final result:
\begin{equation}
\label{breakrate}
\Gamma = \frac{(4N_c)\times 2m}{\pi^2}(\frac{g^2}{4\pi})^2 \exp[-2\pi^2/g^2].
\end{equation}
This is a sensitive function of $g$, varying, for $N_c=3$, from $\Gamma \approx 0.026m$ at $g^2/(4\pi)=1/2$ to $\Gamma \approx 15 m$ at $g^2/(4\pi)=\pi$.  For this large a coupling QCD itself really makes no sense.

\appendix
\section{Appendix:  Conventional Feynman parameters and ``proper-time" Feynman parameters}

The denominator of Fig~\ref{1loop}, with (for example) $N=4$ vertices, has the conventional expression in terms of Feynman parameters $\alpha_j$ associated with the internal lines as labeled:
\begin{equation}
\label{boxden} D=\sum_1^4p_j^2\alpha_{j-1}\alpha_j+(p_2+p_3)^2\alpha_1\alpha_3+(p_3+p_4)^2\alpha_2\alpha_4
\end{equation}
where $\alpha_0\equiv \alpha_4$.  (We ignore irrelevant mass terms in $D$.) Using
\begin{equation}
\label{psum}
p_1^2=-p_1\cdot (p_2+p_3+p_4)
\end{equation}
and three other equations found by permutations one can express the $p_j^2$ in terms of $p_i\cdot p_j,\neq j$.  Then $D$ is
\begin{equation}
\label{deqn}
D=-\sum_{i<j}p_i\cdot p_jx_{ij}(1-x_{ij})
\end{equation}
where $x_{ij}=x_i-x_j$ and 
\begin{equation}
\label{xij}
i<j:\;x_{ij}= \sum^{j-1}_{k=i}\alpha_k 
\end{equation}
(the sum has only the term $\alpha_i$ if $j=i+1$).  We see that $x_{ij}>0$ for $i<j$, and the above expression for $D$ agrees with that found from Eqs.~(\ref{strassres},\ref{bosgreen}).  It is easy to see that Eqs.~(\ref{deqn}, \ref{xij}) hold for all $N$.  OSSIM are equivalent to   the basic expression for a one-loop graph as an integral over a product of propagators $\Delta(k_i)$ in the equivalent of proper-time language by writing each propagator in momentum space as
\begin{equation}
\label{propform}
\Delta (k_i)=\int_0^{\infty}\;\mathrm{d}s_ie^{-s_ik_i^2}
\end{equation}
and changing variables to one overall proper time plus Feynman parameters
\begin{equation}
\label{varchange}
\alpha_i=\frac{\tau_i}{s},\;\sum \alpha_i=1.
\end{equation}
Various orderings of the Feynman parameters $\alpha_i$ correspond to  the $x_i$ introduced above.
For a given ordering of the background lines, ({\em e.g.}, as shown in the labeling of Fig.~\ref{1loop})), one readily finds the factors of the OSSIM expression in Eq.~(\ref{strassres}) except for the last factor, containing information about the interaction with gauge bosons.

\end{document}